# Textual-Based vs. Thinging Machines Conceptual Modeling

**Sabah Al-Fedaghi** *
*Computer Engineering Department*
*Kuwait University*
*Kuwait*
**salfedaghi@yahoo.com, sabah.alfedaghi@ku.edu.kw**

*Abstract – Software engineers typically interpret the domain description in natural language and translate it into a conceptual model. Three approaches are used in this "domain modeling": textual languages, diagrammatic languages, and a mixed based of text and diagrams. According to some researchers, relying on a diagrammatic notation levies certain burdens for designing large models because "visual languages" are intended to depict everything diagrammatically during a development process but fail to do so for a lack of developer efficiency. It is claimed that textual formats enable easier manipulation in editors and tools and facilitate the integration of ontologies in software systems. In this paper, we explore the problem of the relationship between textual format and diagramming in conceptual modeling. The main focus is modeling based on the so-called thinging machine (TM). Several examples are developed in detail to contrast side-by-side targeted domains represented in textual description and TM modeling. A TM model is defined as a thimac (thing/machine) with a time feature that forms dynamic events over static thimacs utilizing five generic actions: create, process, release, transfer, and receive. This provides a conceptual foundation that can be simplified further by eliminating the actions of release, transfer, and receive. A multilevel reduction in the TM diagram's complexity can also be achieved by assuming diagrammatic notations represent the actions of creation and processing. We envision that special tools will help improve developer efficiency. The study's results of contrasting textual and mix-based descriptions vs. TM modeling justify our claim that TM modeling is a more appropriate methodology than other diagrammatic schemes (e.g., UML classes) examined in this paper.*

## I. INTRODUCTION

Conceptual modeling is a significant topic in computer science, impacting areas such as database systems, artificial intelligence, business processes, and software engineering. It is a field that concerns how to specify and develop a conceptual description of a subject domain. Conceptual models are used to support the design of databases, software systems, business processes, enterprises, etc. Diagrammatic languages such as UML, entity relationship models, and other techniques are useful for conceptually understanding a system's purpose, functionality, and requirements before diving into detailed implementation [1].

-----------

Additionally, we human beings think about/understand the world using natural language, scenarios, use cases, narrative sorties, and various other methods, which are effective ways to do so in this context.

Domain modeling is a process of creating a conceptual representation of a portion of a real-world system, often used in software development. It involves identifying key entities and processes and their interconnection in a domain. Three approaches are used for conceptual modeling: textual languages, diagrammatic languages, and combinations of text and diagrams [2]. The latter is usually dominated by diagrams; text is just a supplement.

Conceptual modeling that relies on diagrammatic notation imposes certain burdens in designing large models. Typically, software engineers interpret the problem description in natural language and manually translate it into a model. This "domain modeling" can be time-consuming and highly depends on the expertise of software engineers [3].

### A. Problem: Text-Based vs. Diagrammatic Models

According to [2], experienced developers argue that a text-based conceptual modeling with conventional editors is "much more convenient" in building systems. Text can be printed very easily whereas diagrammatic models, especially for complex and huge systems, often exceed the size of a sheet of paper. In this context, working with graphical tools is time-consuming because modelers must constantly switch between the mouse and the keyboard. This seriously hampers "the creative development process and hinders evolution" [2]. Visual languages are meant to depict everything graphically during a development process, but they fail to do so because of a lack of developer efficiency. If all languages are textual, the integration is much easier and leads to more understandable models [2].

According to [3], in modeling, engineers typically convert a textual domain specification into a domain model represented as a class diagram, and "a domain model uses a subset of class diagram concepts to capture essential elements of a domain, their attributes, and their relationships but does not cover elements related to a more detailed design (e.g., *operations* and *interfaces*)" (italics added). [3] gives an example of a class model and its textual representation, which will be discussed in section IV (Text-Based Modeling).

Conceptual modeling in natural language is a way of modeling semantics, in which texts' semantics are transformed to conceptual models' semantics [4]. According to [5], textual formats enable easier manipulation in editors and tools and facilitate the integration of ontologies in software systems.

In this paper, we explore the problem of the relationship between textual format and diagramming in conceptual modeling. Specifically, we focus on the thinging machine (TM) model. A TM model is defined in terms of TMs with a time feature that forms dynamic events over static thimacs (thing/machines) utilizing five generic actions: create, process, release, transfer, and receive. We analyze the mapping of the textual description of the modeled domain into static, dynamic, and chronology of TM events, and then specify the chronology of TM events in a textual form.

### B. Text

Generally, textual and diagrammatic descriptions complement each other as means of communication in many situations. In software engineering, text-based requirement specification is an active area of research in which natural-language processing tools are used to extract and analyze raw text data. A text-to-model transformation recognizes elements in requirements engineering and generates diagram components from a structured text representation [6].

A text is not merely a series of signals. "Inherent to the text is an internal organization which transforms it on the syntagmatic level into a structural whole" [7]. Structural relations between levels become a specification of the text. Modeling of text must take time into consideration. Time could be seen as events linked to the places they happen [8]. However, mixing static and time aspects is typical in all textual specifications in conceptual modeling. Such an element is important, as we will see with time treatment in TM modeling.

### C. Aims

We claim that TM modeling provides a modeling language that can be applied beyond the strict area of specifying requirements as a first phase of the software development life cycle to create a conceptual model that documents the involved domain in reality. A model is considered a bridge to the design and construction phases. We first attempt to extend this view of TMs to a more general outlook that applies to natural language in the form of "textual knowledge" [9].

### D. Sections

For the sake of a self-contained paper, the following section includes a brief review of the TM model's theoretical framework. Section 3 consists of an example of TM modeling. Section 4 focuses on text-based modeling. Section 5 involves a sample of textual description and informal logic.

## II. TM Modeling

This section includes a summary of the TM model discussed in previous papers (especially recent publications (e.g., [10 and 11])). In contrast to a world comprising substantial things and relationships, as philosophers have overwhelmingly supposed, TM modeling aims at developing a conceptual apparatus for reality based on thimacs. Thimacs are *form*s and *realization*s of processes constructed to include five actions. Their (static) form is a diagrammatic "space" that specifies boundaries and the structure of reality in terms of actions.

### A. Basic Description

The TM model provides us with an ontological representation of reality. It conceptualizes the entities and processes (events) as two modes of one notion, a *thimac*. Thimacs are defined in terms of five actions (Fig. 1). Things, numbers, sets, concepts, and propositions are thimacs.

Thimacs are "discriminative constructs" of the world. The world is divided into thimacs, and various thimacs overlap or combine to form the texture of the whole as a grand thimac. A generic thimac is a gathering of elements into a unity or synthesis of actions: create, process, release, transfer, and receive. The thimac's constituents are formed from the makeup of these actions. An action is a unit of actionality. A TM diagram or sub-diagram is called a *region* at the static-modeling level. Fig. 2 shows an ontological picture that outlines the two levels of the TM scheme.

The synthesis of actions is applied to *events* (at the dynamic level) (see Fig. 2). In a TM, there is no ontological distinction between concrete and conceptual things (e.g., mathematical concepts), and all are things. A thing can be created, processed, released, transferred, and received. The thimac is a machine that creates, processes, releases, transfers, and receives things. All so-called entities, properties, and relationships are thimacs or subthimacs, resulting in a "mesh," of the interconnectedness of all things of a (partial) world with one big complex thimac that contains all within it.

### B. Actions

At the most basic explanatory level, we conceive the thimac as constituted by two kinds of capacity: *thing* and *machine*. A thing constitutes the thimac's capacity to be affected, whether by itself (i.e., one "part" or capacity of the thimac affecting another) or some other thimac (machine). In contrast, the power of machinery is the ability to initiate its actions on other things.

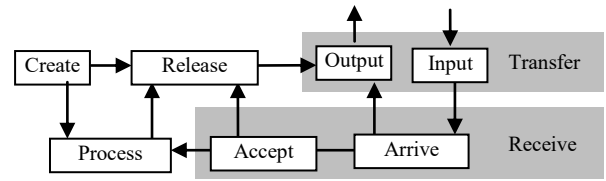

Fig. 1. Thimac.

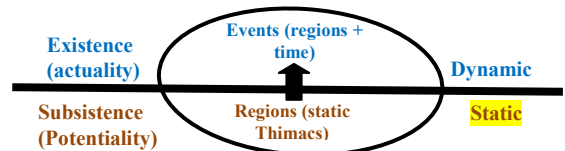

Fig. 2 Two levels of TM modeling.

The thimac is a thing with a "structure" formed by the flow of other thimacs. It is a machine with five actions that operate on things. A thimac's actions, shown in Fig. 1, are described as follows:

*1) Arrive:* A thing arrives at a thimac.

*2) Accept:* A thing enters a thimac. For simplification, the arriving things are assumed to be *accepted* (see Fig. 2); therefore, the *arrive* and *accept* actions are combined into the **receive** action.

*3) Release:* A thing is ready for transfer outside the thimac.

*4) Process:* A thing is changed, handled, and examined, but no new thimac is generated.

*5) Transfer:* A thing crosses a thimac's boundary as input or output.

*6) Create:* Creation here refers to producing something *new to the world* (of the model); therefore, a new thimac is registered as an ontological unit. It indicates the birth or coming-into-thimac (i.e., must make it out of other thimacs). At the static level, *creation* is a logical possibility of realization at the dynamic level.

The TM diagrammatic model includes *storage* (represented as a **cylinder** in the TM diagram) and *triggering* (denoted by a **dashed arrow**). Triggering transforms from one series of movements to another (e.g., electricity triggers heat generation).

### III. EXAMPLE

According to [12], the critical use case work involves writing text, not diagrams, or focusing on relationships. Spending time working on a use case diagram and discussing use case relationships rather than focusing on writing text misplaces the relative effort.

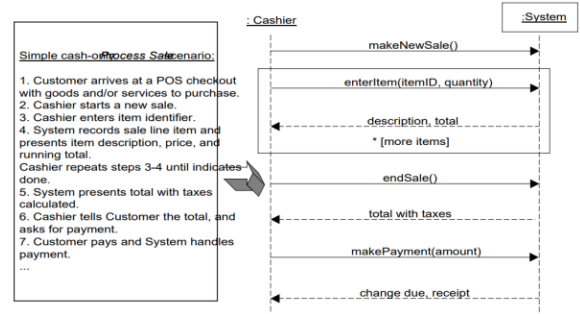

Fig. 3 Sales sequence diagram (From [12]).

However, according to [12], system behavior is a description of what a system does without explaining how it does it. One part of that description is a system sequence diagram. A sequence diagram shows the *events* that external actors generate in their order and inter-system events. [12] then gives an example for the process sale use case. It indicates that the cashier generates the *makeNewSale*, *enteritem*, *endSale*, and *makePayment* events (see Fig. 3).

#### A. Static Model

Fig. 4 shows the TM static model.

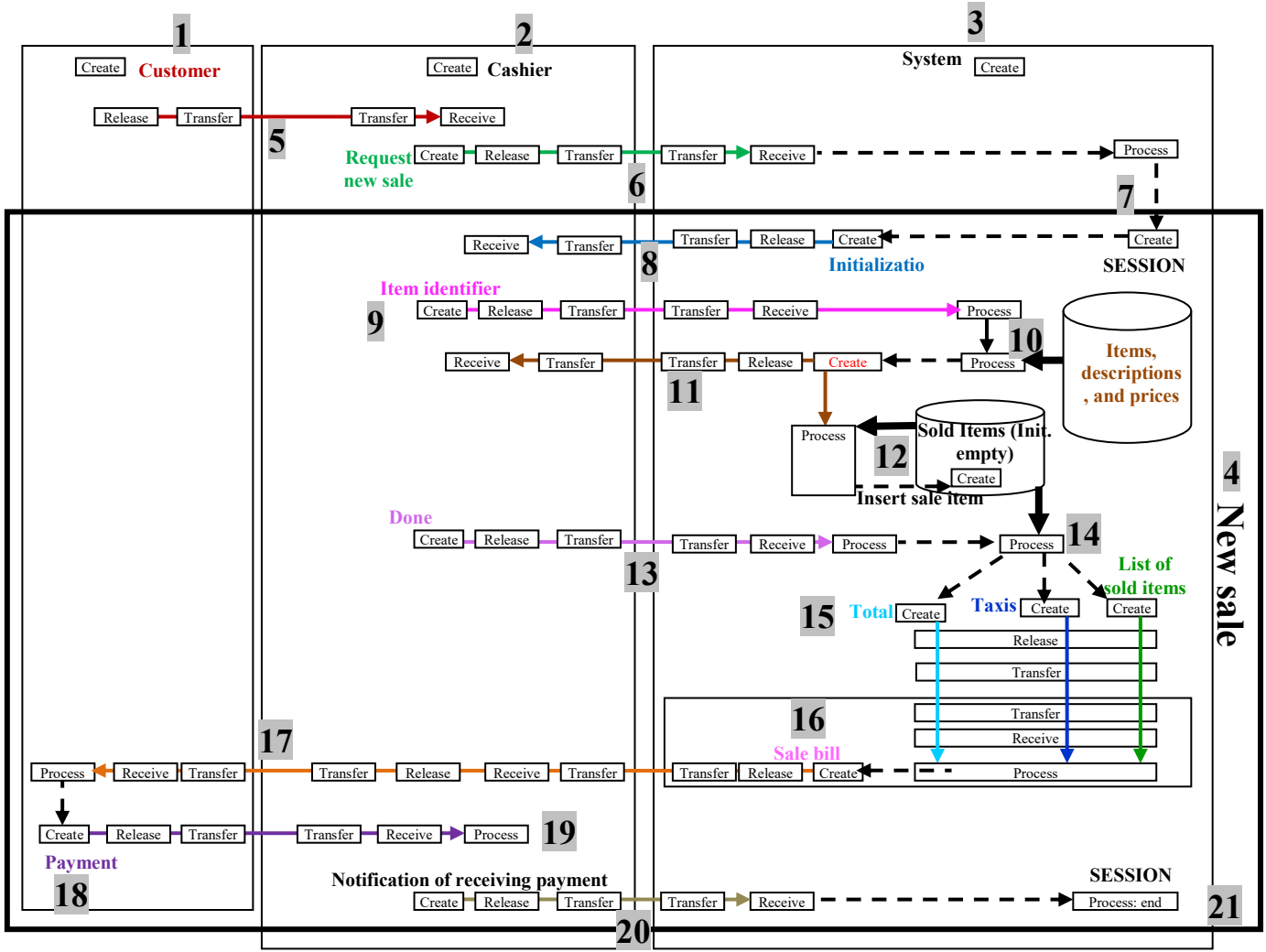

Fig. 4 The static model.

There are three main thimacs: User (1), Casher (2), and System (3) crossed by the thimac new sale (4). Accordingly, the model can be described as follows:

- The customer goes to the checkout with goods and/or services to purchase (5).
- The cashier requests to start a new sale (6).
- The system receives the request, initiates a new sale session (7), and replies to the cashier request (8).
- The cashier inputs an item identifier that flows to the system (9).
- The system finds the description and price of the input item (10) and sends them to the cashier (11). Note that extracting the description and price involves, for simplicity's sake, mixed types of operation of an identifier value and a file that contains all descriptions and prices. It is possible to model the exact operation, say, retrieval sequentially and comparison with the identifier. Additionally, note that the *creation* of the description and price indicates the appearance of the description and price after being hidden inside the file.
- The sale details (identifier, description, and price) are also added to the file of the sold items (12). We assume that this file is initially empty.
- At the end, the cashier sends the "done" message, indicating the end of the sale for the customer (13).
- The system retrieves elements in the sold-items file (14), extracts the items list, and calculates the total and taxes (15), thus creating the final sale bill (16), which is sent to the cashier, who shows it to the customer (17).
- The customer pays (18), and the cashier receives and processes the payment (19).
- The cashier informs the system about the payment (20), and the system finishes the sale session (21).

If the arrows sufficiently indicate the flow's direction, Fig. 4 can be simplified by deleting release, transfer, and receive. Fig. 5 shows this simplification.

### B. Dynamic Model

Fig. 6 shows the dynamic model for the example using the simplified static model. It includes the following six events.

$E_1$: The cashier requests and triggers the creation of a new session.

$E_2$: The system starts a new sales session.

$E_3$: The system creates an initial screen for the cashier.

$E_4$: The cashier inputs an item's identity, and the system finds its description and price and shows them to the cashier.

$E_5$: The cashier sends a "done" signal. The system calculates the total and taxes and sends the final bill to the cashier.

$E_6$: The cashier seems to signal that the payment has been received. Accordingly, the system ends the session.

Fig. 7 shows the corresponding chronology of events. Fig. 8 shows an alternative "carving" of events, probably more suitable for software engineers.

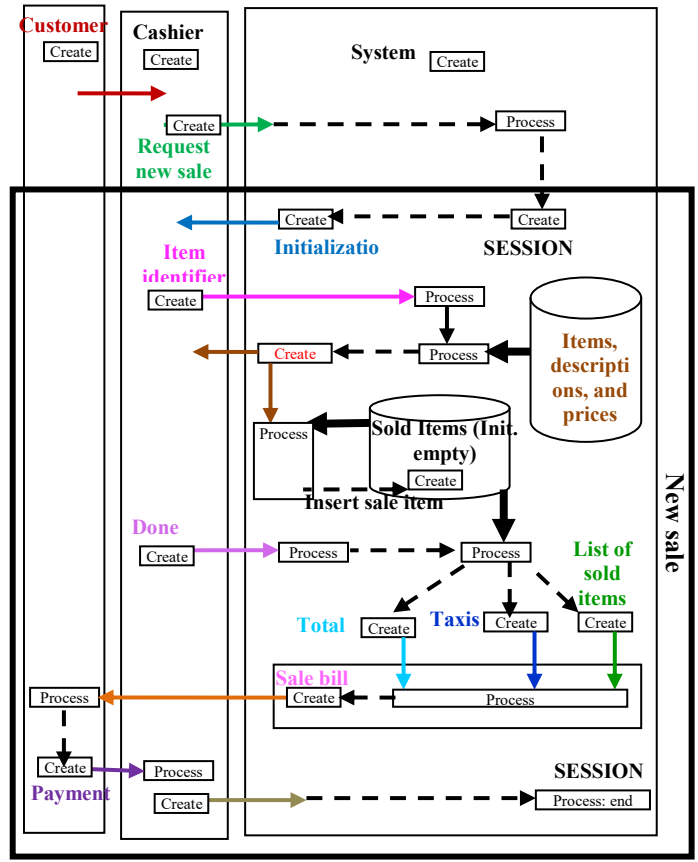

Fig. 5 The simplified static model.

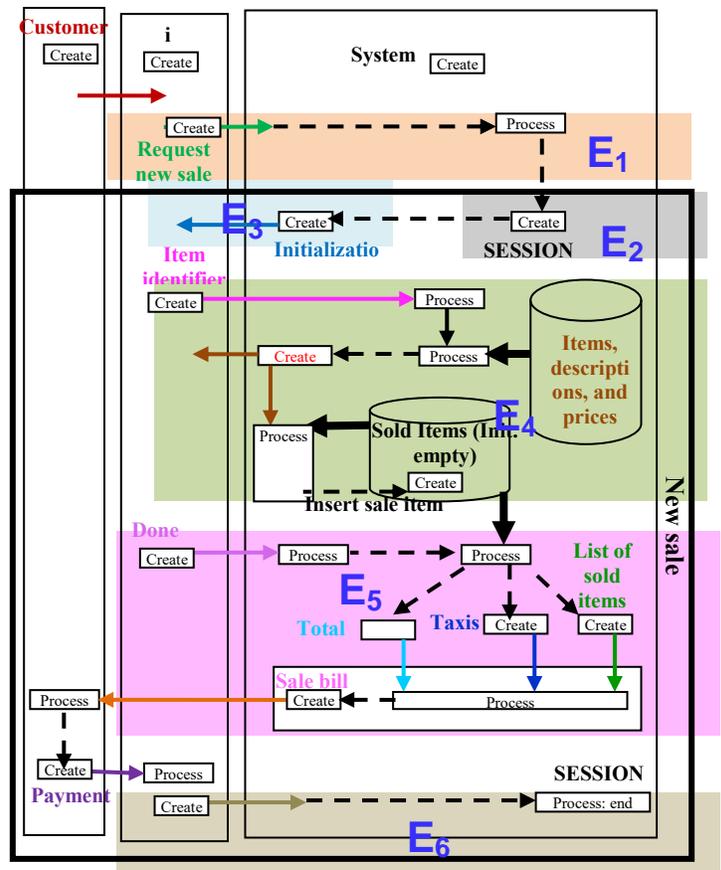

Fig. 6 The dynamic model.

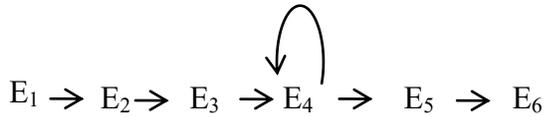

Fig. 7 Chronology of events.

We can note that the diagrammatic TM modeling potentially refines and focuses the textual description for various purposes.

## IV. TEXT-BASED MODELING

In this section, we show that the TM model facilitates alternative textual/diagrammatic views that can be provided to the modeler. These views include the static, the dynamic, and the chronology of events used to develop a textual specification.

According to [3], modeling engineers typically convert a textual domain specification into a domain model represented as a class diagram, and "a domain model uses a subset of class diagram concepts to capture essential elements of a domain, their attributes, and their relationships but does not cover elements related to a more detailed design (e.g., operations and interfaces)." [3] gives the example of a Helping Hand Store (H2S) in terms of the class model (Fig. 9) and its textual representation.

### A. Static Model

Fig. 10 shows the H2S TM static model with some assumed simplifications, described as follows:
- The volunteer (gray number 1) updates the system with his/her request to deliver (2) requirements (e.g., available times) that flow to the system (3) to be processed (4) to create an address list (5) that is not sent to him/her. The address list includes items to be collected from residents or delivered to the client.
- The volunteer enters the assigned vehicle (7) and drives to either the resident (8) or the distribution center (9), depending on his/her assignment.
- When arriving at the residence (10 and 11), they pick up donated items (12 and 13) and go to the distribution center (14, 15). Note that the loop of visiting several residents will be specified in the chronology of events, which will be determined later.
- The volunteer delivers the picked up items (16 and 17) to the distribution center. Note that "create" at (17) refers to the item's first appearance in the distribution center.
- Alternatively, the volunteer (in their vehicle) goes to the distribution center (9) to pick up items (18) and take them to clients (19 and 20).
- The resident updates their list of available items to donate and their address (21, 22, and 23).

### B. Dynamic Model

Fig. 11 shows the dynamic model in which events are "carved" from the static model.

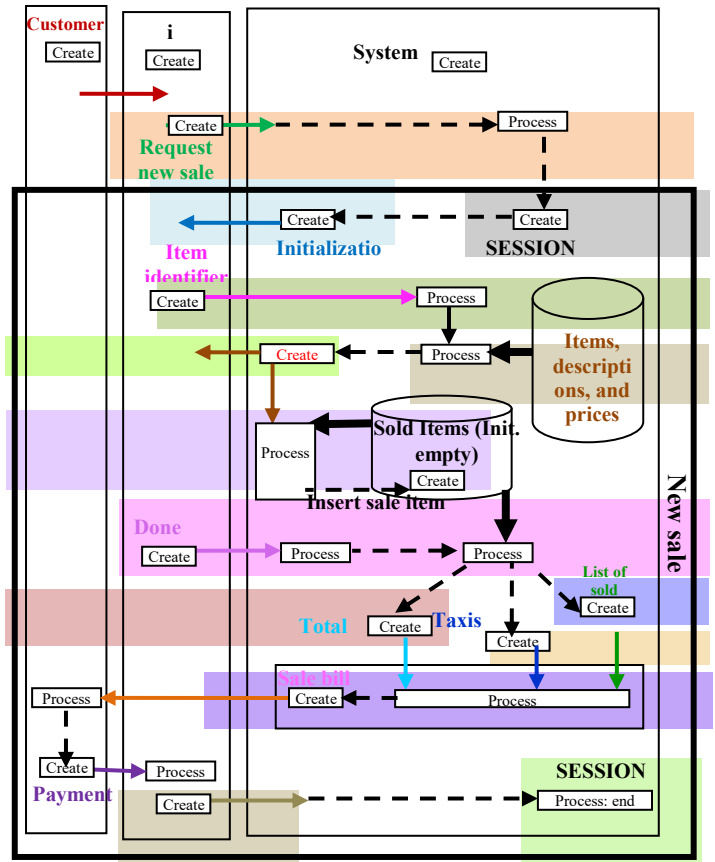

Fig. 8 Alternative dynamic model.

**Textual Domain Description:** The Helping Hand Store (H2S) collects secondhand articles and nonperishable foods from residents of the city and distributes them to those in need. H2S operates in many cities, but each location is run independently. To increase the number of items available for distribution, H2S is seeking to offer a Pickup and Delivery Service to its customers, which would allow a resident to schedule a pickup of items from a street address online at the H2S website. A resident enters a name, street address, phone number, optional email address, as well as a description of the items to be picked up. H2S has a fleet of pickup vehicles, which it uses to collect items from residents. A pickup route for that day is determined for each vehicle for which a volunteer driver is available. Volunteer drivers indicate their available days on the H2S website. The route considers the available storage space of a vehicle and the dimensions and weights of scheduled items. A scheduled pickup may occur anytime between 8:00 and 14:00. After completing all scheduled pickups, the driver drops off all collected secondhand articles at H2S's distribution center. Non-perishable foods, on the other hand, are directly dropped off at the food bank, which then deals with these items without further involvement from H2S. (From

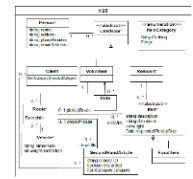

Fig. 9 H2S textual description and class diagram (From [3])

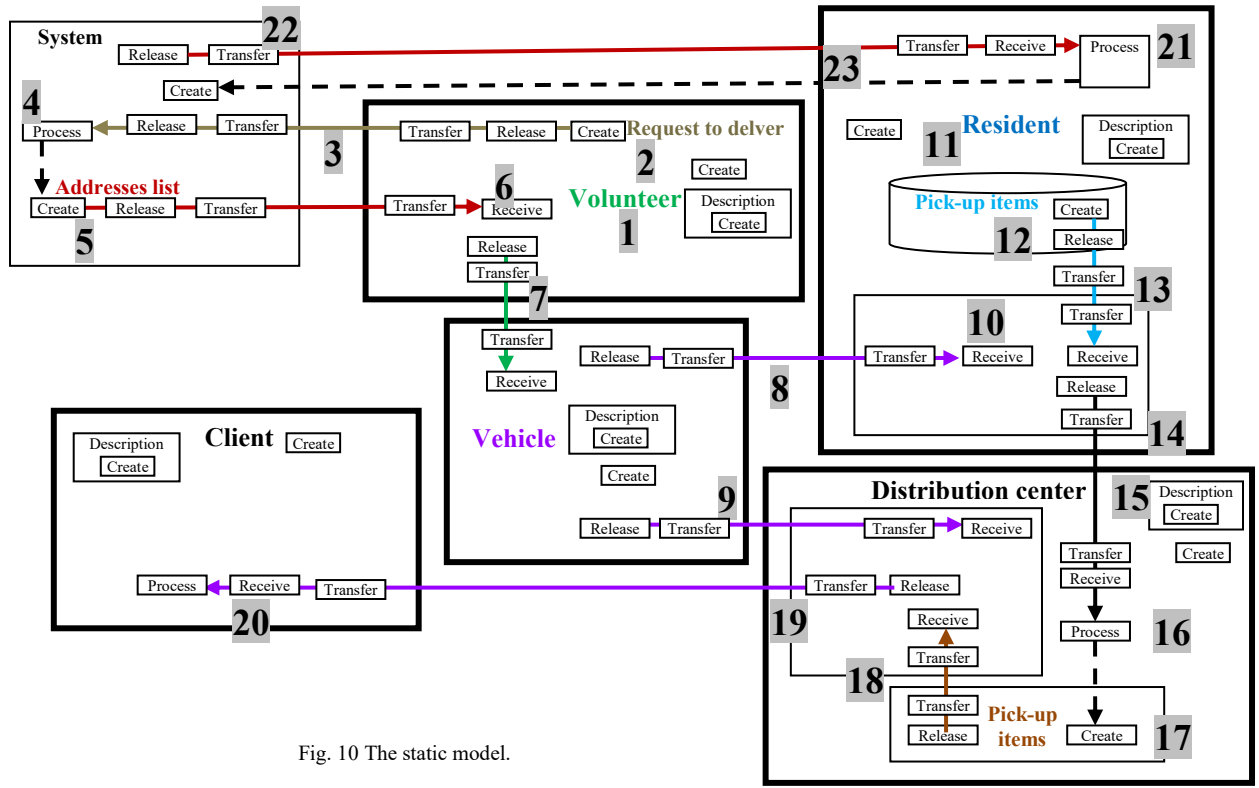

Fig. 10 The static model.

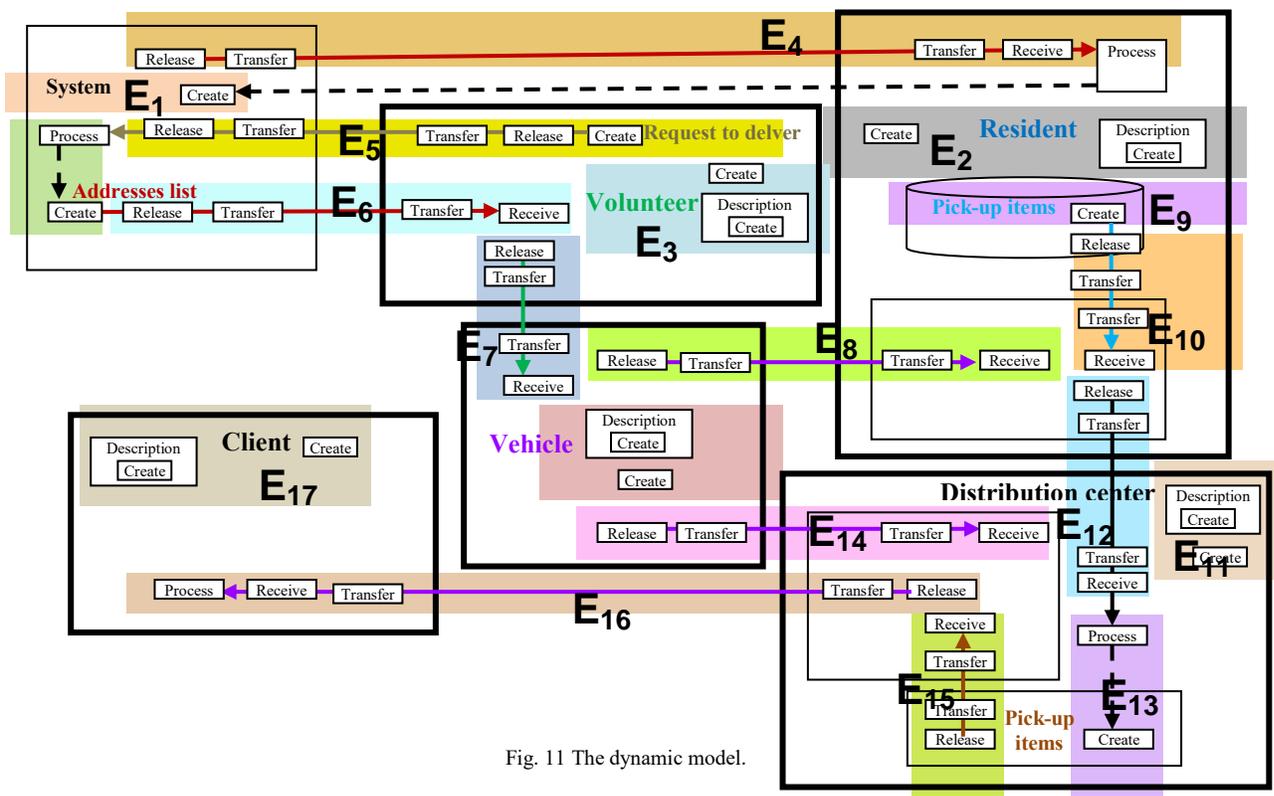

Fig. 11 The dynamic model.

$E_1$: System is active.
$E_2$: A resident is present.
$E_3$: A volunteer is present.
$E_4$: A resident updates the system (e.g., new items to pick up).
$E_5$: A volunteer requests to deliver an item.
$E_6$: Address list sent to the volunteer.
$E_7$: The volunteer accesses a vehicle.
$E_8$: The volunteer goes to a resident.
$E_9$: Pickup items are available.
$E_{10}$: The volunteer picks up items.
$E_{11}$: The distribution center is open.
$E_{12}$: The volunteer drives to the distribution center.
$E_{13}$: The volunteer delivers the items to the center.
$E_{14}$: The volunteer goes to the distribution center.
$E_{15}$: The volunteer picks up the items from the vendor.
$E_{16}$: The volunteer delivers the items.
$E_{17}$: The client is present.

Fig. 12 shows the chronology of events for this example.

We claim the TM model provides a static and dynamic picture of the involved domain, in comparison to the textual description and the accompanied class diagram. We cannot substantiate this claim except by putting the approaches side-by-side to contrast them.

Fig. 13 shows the textual description of the chronology of events. Incorporating the natural-language processing technique (e.g., [13]) in analyzing the transformation between the TM model and various versions of the problem text is a very interesting research problem.

Fig. 14 shows the parts of the original text modeled in the TM. An analysis can modify the TM model by incorporating missing parts (e.g., picking up between 8:00 and 14:00). Other portions may not be necessary details (e.g., H2S operates in many cities, but each location runs independently). However, we can see that the TM model facilitates mapping and alternative textual/diagrammatic views that can be provided to the modeler. The static, dynamic, and chronology of events and their textual specification are realigned systematically as multiple versions of the same modeled domain.

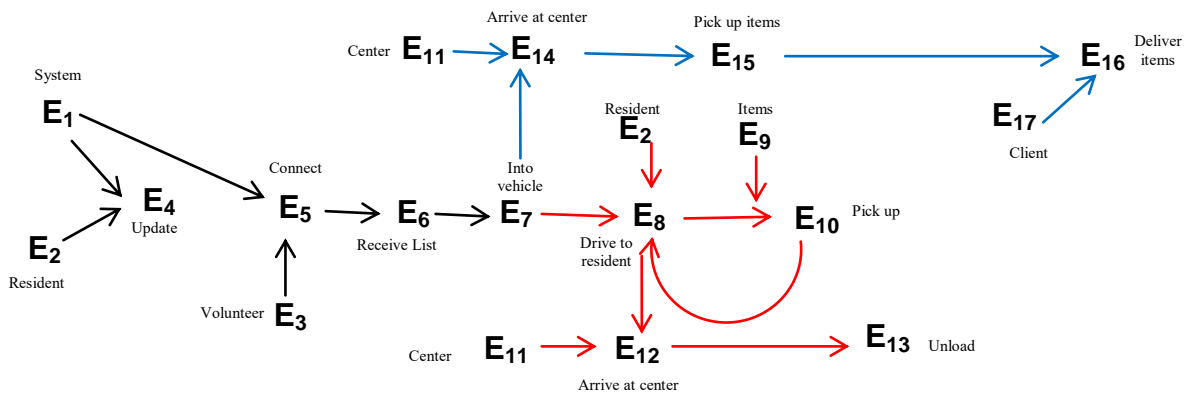

Fig. 12 Chronology of events

The system is active ($E_1$), and a resident is present ($E_2$). A resident updates the system (e.g., new items to pick up) ($E_4$). Since $E_1$, a volunteer is present ($E_3$). He/she requests delivering item ($E_5$). **Accordingly**, the address is list sent to the volunteer ($E_6$). The volunteer accesses a vehicle ($E_7$). **Then, either** ($E_{16}$)

(a) The volunteer goes to a resident ($E_8$) **and**, assuming pickup items are available ($E_9$), he/she picks up items ($E_{10}$), **repeating this for all residents in the list**. **At the end**, **and** assuming that the distribution center is open ($E_{11}$), the volunteer drives to the distribution center ($E_{12}$) **to** deliver the items ($E_{13}$).

(b) The volunteer goes to the distribution center ($E_{14}$) **to** pick up the items from the center ($E_{15}$), **and** he/she delivers the items ($E_{16}$) to the client ($E_{17}$).

Fig. 13 Textual description of the chronology of events

**Textual Domain Description: The Helping Hand Store (H2S)** collects secondhand articles and nonperishable foods from city residents and distributes them to those in need. H2S operates in many cities, but each location is run independently. To increase the number of items available for distribution, H2S is seeking to offer its customers a pickup and delivery service, which would allow residents to schedule a pickup of items from a street address on the H2S website. A resident enters a name, street address, phone number, optional email address, and a description of the items to be picked up. H2S has a fleet of pickup vehicles, which it uses to collect items from residents. A pickup route for that day is determined for each vehicle for which a volunteer driver is available. Volunteer drivers indicate their available days on the H2S website. The route considers the vehicle's available storage space and the dimensions and weights of scheduled items. A scheduled pickup may occur anytime between 8:00 and 14:00. After completing all scheduled pickups, the driver drops off all collected secondhand articles at H2S's distribution center. Nonperishable foods, on the other hand, are directly dropped off at the food bank, which then deals with these items without further involvement from H2S.

Fig. 14 The red text indicates events already modeled in the TM.

## V. TEXT AND INFORMAL LOGIC

According to [14], argument diagramming is a widely used technique in informal logic. The technique represents the reasoning structure in a given argument found in a text. The whole network of points and lines represents a kind of overview of the reasoning in the given argument, showing the various premises and conclusions in the chain of reasoning.

In such a technique, the user begins constructing a diagram by inserting the text of the argument into a text document. The next step is to identify each statement that is a premise or a conclusion in the argument by highlighting it. By this means, an argument diagram is constructed of the kind illustrated in Fig. 16.

### A. Static Model

Fig. 17 shows the TM static diagram that models the text in Fig. 15 described as follows.

- You (3) have considered (2) your weight and how to reduce it (4).
- Enough milk (5) in your reduced-calorie diet (6) could provide the nutritional support (7) you need for healthy, effective weight loss (8).
- Emerging research (9) shows that drinking three glasses of milk daily (10) when dieting (11) may result (12) in the loss of body fat (13) while maintaining muscle (14).
- The calcium and protein in milk (15 and 16) may help explain (17) these weight loss benefits (18). Recent studies (19) indicate that calcium (15) is part of the body's natural system for burning fat (20) and protein is essential for building and keeping muscle (21).
- Milk is the only beverage (22) that naturally provides the unique combination (23) of calcium and protein for healthy, practical weight loss support (24).
- No other single food item (25) provides more calcium in the United States diet (26) than milk.
- It's time to add healthy weight loss (8) to the already extensive list of good things (27) that milk can do (28) for your body (29).
- If you're serious (30) about losing weight the healthy way (8), make sure to exercise (31), limit (32) your calories (33), and drink at least three glasses (10) a day of low-fat or fat-free milk (34).

### B. Dynamic Model

We select the set of events as follows (see Fig. 18).

$E_1$: You exist.
$E_2$: You consider losing weight.
$E_3$: Include enough milk in your reduced-calorie diet.
$E_4$: Providing the nutritional support you need.
$E_5$: Providing healthy, effective weight loss.

Looking to drop a few pounds (of **weight**)?
Including **enough milk** in your **reduced-calorie diet** could provide the **nutritional support** you need for **healthy, effective weight loss**. In fact, emerging research suggests that drinking three glasses of milk daily when dieting may promote the loss of body fat while maintaining muscle.
The calcium and protein in milk may help explain these weight loss benefits. Recent studies indicate that calcium is part of the body's natural system for burning fat and protein is essential for building and keeping muscle.
And milk is the only beverage that naturally provides the unique combination of calcium in protein for healthy, effective weight loss support. In fact, no other single food item provides more calcium in the U.S. diet than milk. It's time to add healthy weight loss to the already extensive list of good things that milk can do for your body.
If you're serious about losing weight the healthy way, make sure to exercise, limit your calories, and drink at least three glasses a day of low fat or fat-free milk, which has the same amount of calcium, protein, and other nutrients as whole milk.

Fig. 15 The text of the argument (from [14]).

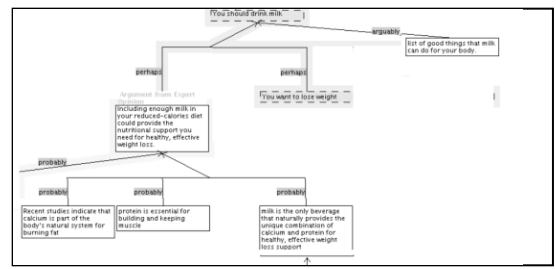

Fig. 16 Diagram for the milk argument (partial – from [14]).

$E_6$: There is emerging research.
$E_7$: Drinking three glasses of milk when dieting.
$E_8$: Loss of body fat while maintaining more muscle.
$E_9$: Maintaining muscle.
$E_{10}$: Loss of body fat.
$E_{11}$: There is calcium in the milk.
$E_{12}$: There is protein in the milk.
$E_{13}$: There is a combination of calcium and protein in the milk.
$E_{14}$: There is an explanation
$E_{15}$: There is recent research.
$E_{16}$: Calcium is part of the body's natural system for burning fat.
$E_{17}$: Protein is essential for building and keeping muscle
$E_{18}$: There is a unique (only) beverage.
$E_{19}$: No other single food ('negative' event see [10] and [11])).
$E_{20}$: There is more calcium.
$E_{21}$: There is a U.S. diet.
$E_{22}$: There is a list of good things that milk can do for your body.
$E_{23}$: You are serious.
$E_{24}$: You exercise.
$E_{25}$: You limit your calories.
$E_{26}$: You drink at least three glasses of low-fat or fat-free milk daily.

Fig. 19 shows the chronology of events for the milk example. Fig. 20 illustrates the chronology of events through the grouping of events.

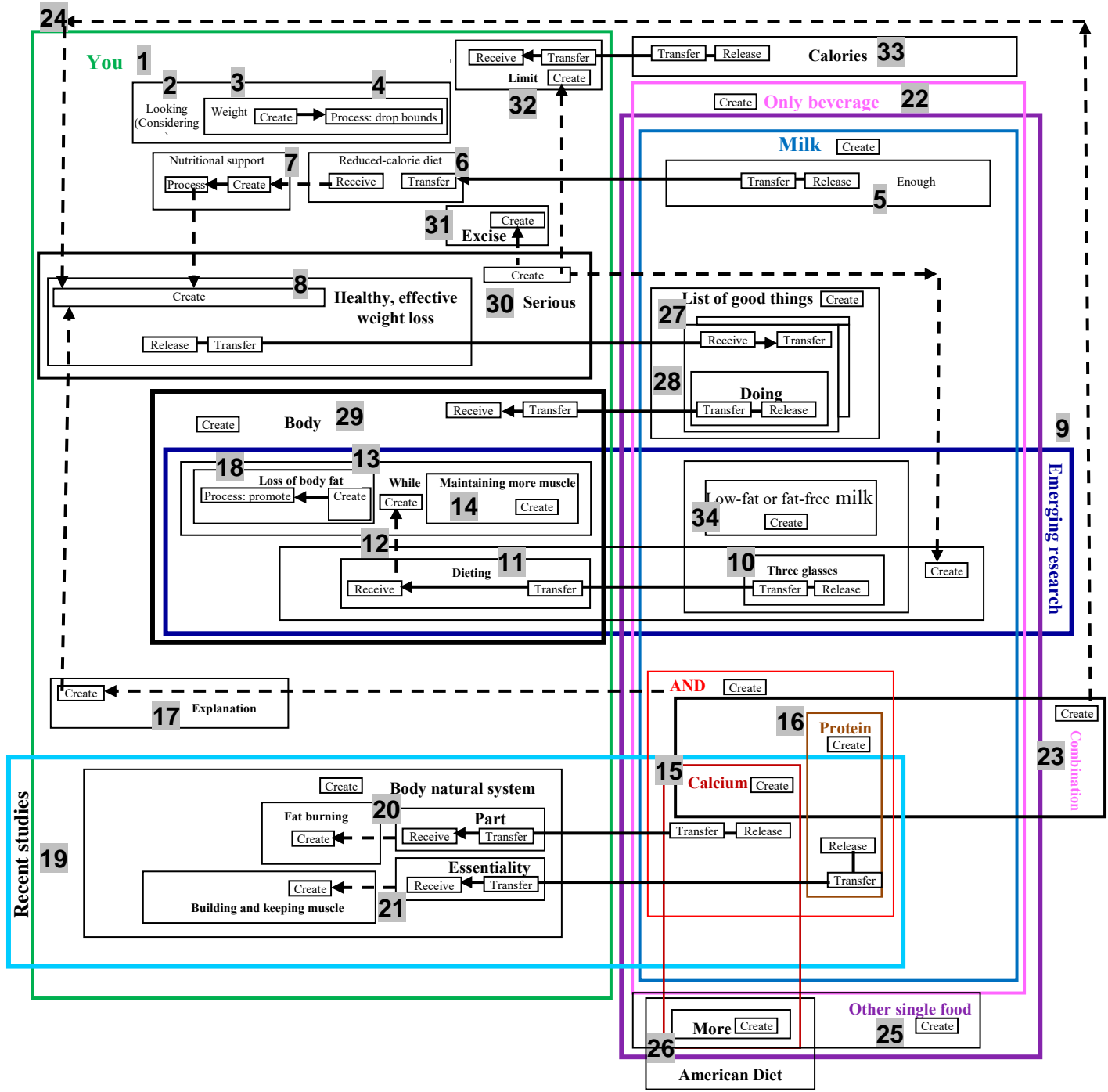

Fig. 17 Static TM model.

Accordingly, we can organize the informal argument as follows (note the colors in Fig. 20).

You want to lose weight? Include milk in your reduced-calorie diet because
**1- It provides nutritional support for weight loss.**
**2- It is a unique beverage for nutritional support for weight loss.**
3- Drinking three glasses daily helps you maintain muscle while losing fat.
4- The calcium and protein in milk help explain its weight loss benefits because calcium is part of the body's natural system for burning fat and protein is essential for building and keeping muscle.
5- No other single food item provides more calcium to the U.S. diet than milk.
Therefore, if you are serious about losing weight, then exercise, limit your calories, and drink at least three glasses a day of low-fat or fat-free milk

Fig. 18 Dynamic TM model.

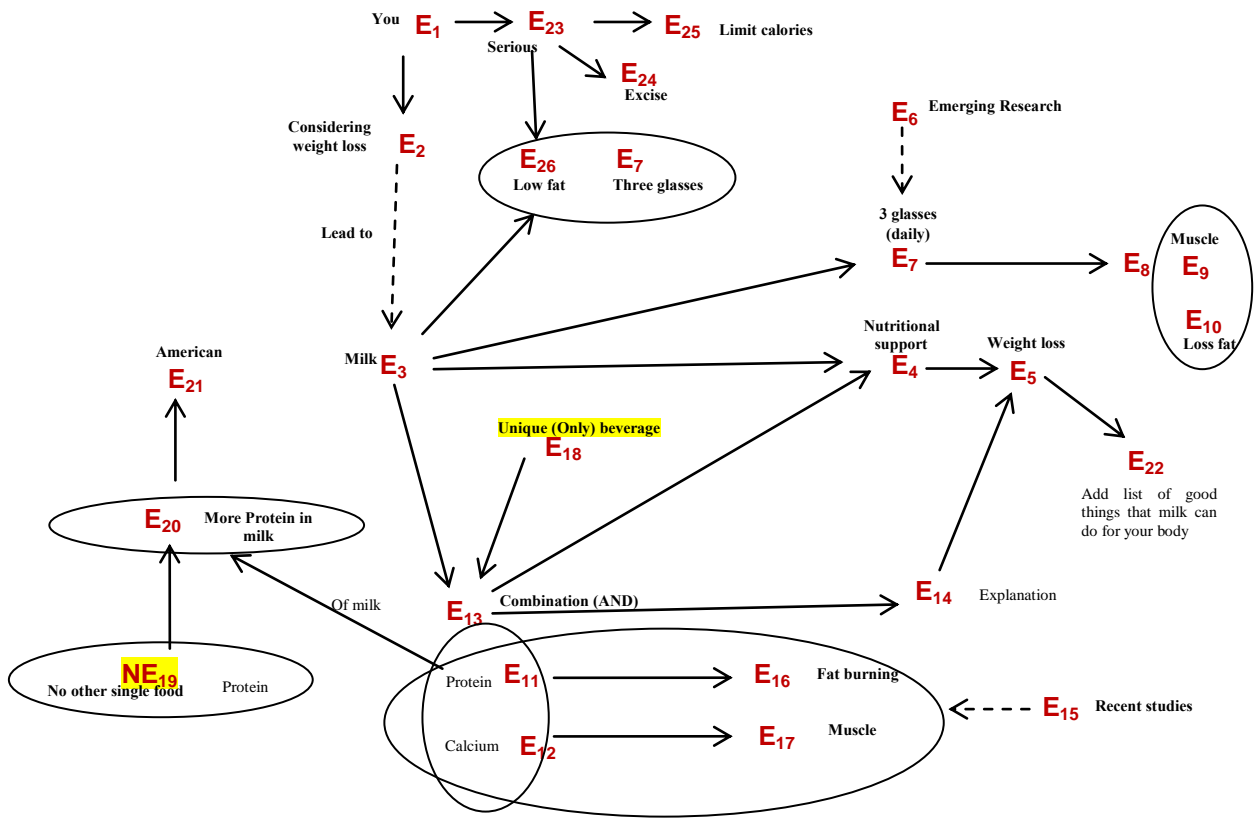

**Fig. 19 Chronology of events**

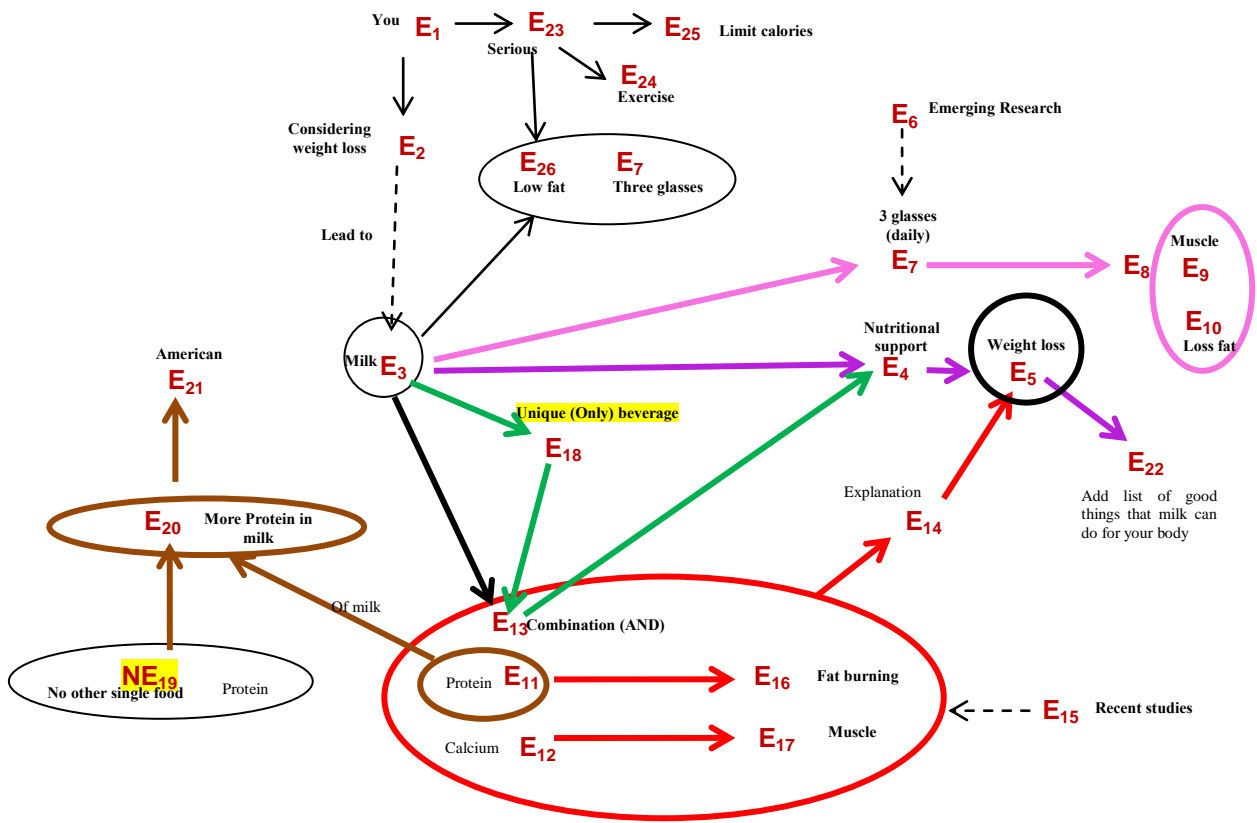

Fig. 20 Illustrative chronology of events.

## CONCLUSION

In this paper, we have explored the problem of the relationship between textual format and diagramming in conceptual modeling. The main focus is on modeling based on TMs. Several examples are developed in detail to contrast side-by-side targeted domains represented in textual description and TM modeling. We show that a TM model is uniquely well founded on an ontological scheme that includes static and dynamic levels, utilizing five generic actions: create, process, release, transfer, and receive. This is contrasted with the mixed textual representation of staticity and dynamism. Additionally, the limited number of actions in TMs gives the semantics of the intended domain a more restrictive form, thus limiting the possibility of ambiguity.

Although the research on the relationship between textual description and TM modeling is still too early to characterize fully, the material in this paper points to improvement in representing meanings among concepts embedded in natural language texts. TM modeling can be used in many applications to enforce structure over textual entities and processes. The structure of something is how its parts are put together. Therefore, the structure of a text (e.g., words, sentences, paragraphs) is mapped to the structure of (multilevel) thimacs formed from actions. The purpose is to capture the key components and their interconnections. The examples in this paper show that this is an achievable goal. Some issues remain (e.g., representing *may be*, *could be*) and need further thought.

A TM of a certain text would help conceptual modelers exert more control over the text content, for they can filter and revise the embedded information in a repetitive process of improving the diagrammatic TM representation.